\documentclass{article}
\usepackage{arxiv}
\usepackage[utf8]{inputenc} 
\usepackage[T1]{fontenc}    
\usepackage{hyperref}       
\usepackage{url}            
\usepackage{booktabs}       
\usepackage{amsfonts}       
\usepackage{nicefrac}       
\usepackage{microtype}      
\usepackage{lipsum}		
\usepackage{graphicx}
\usepackage{natbib}
\usepackage{doi}

\title{Predictability of Global AI Weather Models}

\author{\href{https://orcid.org/0000-0001-8947-8534}{\includegraphics[scale=0.07]{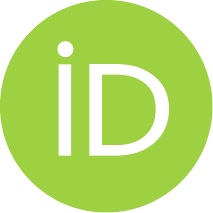}\hspace{1mm}Chanh Kieu}\thanks{.} \\
	Department of Earth and Atmospheric Sciences\\
	Indiana University Bloomington, 47405, Indiana\\
	\texttt{ckieu@iu.edu} \\
}

\renewcommand{\shorttitle}{\textit{arXiv} AI weather predictability}

\hypersetup{
pdftitle={A template for the arxiv style},
pdfsubject={q-bio.NC, q-bio.QM},
pdfauthor={David S.~Hippocampus, Elias D.~Striatum},
pdfkeywords={First keyword, Second keyword, More},
}

\begin{document}
\maketitle

\begin{abstract}
This study examines the predictability of artificial intelligence (AI) models for weather prediction. Using a simple deep-learning architecture based on convolutional long short-term memory and the ERA5 data for training, we show that different time-stepping techniques can have a strong influence on the model performance and weather predictability. Specifically, a small-step approach for which the future state is predicted by recursively iterating an AI model over a small time increment displays strong sensitivity to the type of input channels, the number of data frames, or forecast lead times. In contrast, a big-step approach for which a current state is directly projected to a future state at each corresponding lead time provides much better forecast skill and a longer predictability range. In particular, the big-step approach is very resilient to different input channels, or data frames. In this regard, our results present a different method for implementing global AI models for weather prediction, which can optimize the model performance even with minimum input channels or data frames. Our method based on the big-step approach can be also extended to search for a practical predictability limit in any chaotic system. 
\end{abstract}


%
%
\section{Introduction}\label{sec:introduction}
Numerical weather prediction (NWP) models are built on a set of governing equations, which dictates all physical and dynamical processes in the atmosphere \citep{ECMWF2023, Holton2004, Kalnay2002}. The traditional approach for all NWP models is based on finite or spectral numerical methods, which propagate information from an initial condition with time at a given time step increment. So long as the time step for these numerical schemes can ensure certain stability constraints, the model can be integrated forward to predict a new state of the atmosphere from any given initial and boundary condition consecutively.

Due to the chaotic nature of the atmosphere, all current NWP model integration cannot be taken forever long. Depending on the weather patterns, variables, and forecast metrics, the accuracy of NWP models gradually decays after about 2 weeks, justifying the commonly known as a 2-week limit in weather forecast \citep{DelSoleTippett2007, ECMWF2023, Kalnay2002, Leith1971, Lorenz1982, Lorenz1990, Shukla}. Operationally, this decay of model accuracy is often examined in terms of 500 hPa geopotential height errors, which show an error saturation after about $\approx$ 2 weeks beyond which model prediction is not better than a simple climatological or persistent forecast \citep{Krishnamurthy2019, Lorenz1982, Shukla, Palmer_etal2014, Wattmeyer_etal2021, Weyn_etal2021}. Such a limited range underlines the key concept of atmospheric predictability, which was laid out formally in a series of early studies by Lorenz and many others \citep{Lorenz1969, Lorenz1982, Lorenz1990, metais_and_marcel1986, Shukla}. 

In practical applications, atmospheric predictability is not a universal concept, but it depends on the specific metric one examines as well as the time scale used to define the reference climatology \citep{DelSole_etal2017}. In fact, the 2-week predictability limit for global NWP models is traditionally based on 500-hPa geopotential height. For other metrics such as rainfall, tropical cyclone intensity, tornadoes, or convective-scale systems, their predictability is different and must be re-defined for each variable, depending on the nature of each weather system and reference climatology. For example, tropical cyclone intensity has a much shorter predictability range of 1-3 days after reaching the maximum intensity stage \citep{KieuMoon2016, kieu_etal2018}. Likewise, convective- to meso-scale precipitation has a similarly short range of 1-2 days instead of 2 weeks as obtained from global-scale 500-hPa geopotential height \citep{Weisman_etal2008, Islam_etal93}.   

Given the inherent chaotic nature of the atmospheric governing equations, a factor that can largely affect the estimation of atmospheric predictability in an NWP model is the way the model is numerically integrated with time. Different time discretization schemes can accumulate and/or amplify different types of numerical errors, depending on the stability of these schemes \citep{ECMWF2023, Kalnay2002}. As such, any small initial or model errors can rapidly grow and spread across scales at different rates, leading to the deterioration of model accuracy with forecast lead times \citep{Lorenz1992, Palmer_etal2014}. These model internal errors as well as initial condition uncertainties are confounding and generally hard to separate from the practical perspective. Their combined effects contribute to an error saturation in model forecasts and account for limited weather predictability in any NWP model.       

Recent developments in AI present a very different pathway to weather prediction \citep{Bi_etal2022, Lam_etal2023, Nguyen_etal2023, Pathak_etal2022, Weyn_etal2019}. Building on data-driven models that can learn rules from a vast amount of input data, AI models search for a set of parameters (weights) that minimize the difference between model predictions and ground truths via a loss function. Given a training dataset and a model architecture, one can technically train a model that relates two weather states at times $t$ and $t+\Delta$, where $\Delta$ is the model time step. By design, this model then allows one to predict a new state at any lead time $T = n \Delta$ by iterating the model recursively every increment $\Delta$, which underlines all current AI weather prediction models.      

A particular question arisen for these AI weather models is \textit{"can these data-driven models realize the limited predictability of the atmosphere as established in physical-based models?"}. This is an important question, because our ability to develop and successfully train an AI model would depend on the answer to this question. In fact, the existence of weather chaos is the key factor that distinguishes the applications of AI models for weather prediction from traditional computer vision or image processing. For the latter, a good AI model should correctly detect, understand, or classify an object regardless of the quality of an input image. For weather prediction, a good AI model should contain chaos as a part of its properties, which however prevents one from predicting weather after a certain lead time when input data contains some initial uncertainty \citep{Kieu2024}. This is a dilemma for all AI weather models, because, on the one hand, one tries to train the best possible model for weather prediction, which must include chaos as an intrinsic property of atmospheric dynamics. On the other hand, this same atmospheric chaos will quickly destroy the causal relationship between an initial condition and a future state due to the unavoidable existence of noise in the initial condition, thus preventing one from training any AI model for long lead times. 



In this study, we will explore the predictability of AI models for weather forecasting, using a simplified global deep-learning model. Our specific aim is to quantify the predictability for an AI model and examine how it varies with different time-stepping approaches, given the chaotic nature of weather systems. While the time-stepping technique is just one aspect of AI models that can affect predictability, this aim is still very ambitious. This is because fully addressing this issue would require the construction of the best possible AI models as well as training datasets for weather prediction, which are beyond what we can afford. Within the scope of this report, we will however use a simple convolutional deep learning architecture to demonstrate our main objective and interpretation. More extensive analyses of global weather predictability for different metrics/variables can be readily followed by using the same methodology as presented herein, for which we will report in an upcoming full article.
%
%
\section{Global AI weather model}\label{sec:model}
Given the nature of weather prediction based on data-driven models, it is important to choose an architecture that could effectively learn the rules of atmospheric evolution from input data. There are many different architectures that could fit this purpose. Here in this study, we choose a simple deep-learning model that is built on convolutional LSTM (\texttt{ConvLSTM}) blocks, which are available in the Tensorflow-supported Keras library. This choice is motivated by the fact that the large-scale evolution of the atmosphere must inherit (or memorize) some previous states of the atmosphere due to its large inertia. With climate data often provided at a fixed time interval on a regular grid, it is thus natural to choose \texttt{ConvLSTM} that can be optimized for such a spatiotemporal data structure.   

Following the same architecture proposed by \cite{Shi_etal2015}, we designed our global AI weather prediction (AWP) model with 4 \texttt{ConvLSTM} layers, which have 32, 64, 128, and 256 hidden states, respectively. For the last layers, note that we use a two-dimensional convolutional block with the same number of filters as the number of input channels, which ensures that the output tensor will have the same size as the input tensor. Instead of a three-dimensional convolutional block as in Shi et al, our two-dimensional (2D) convolutional block is applied to each frame independently, because the 3D convolution with \texttt{padding=same} in Keras would introduce a bogus zero padding to the time dimension that may distort the evolution of the weather state at a long lead time to be described in Section \ref{sec:description}. Thus, the two-dimensional convolution layer is used for our last layer here, which indeed improves the performance of our AWP model significantly at long forecast lead times.  

%
%
\begin{table}[ht]
\centering
\caption{The AWP model design based on the Keras \texttt{ConvLSTM} architecture used for the control experiment \texttt{Exp1} in this study (see Table \ref{tab:experiments} for a list of other experiments.}
\label{tab:model}
\scalebox{1.1}{
\begin{tabular}{lll}
\hline
Layer (type) & Output Shape & Param \# \\
\hline
input\_1 (InputLayer) & [(None, 5, 32, 64, 5)]  &  0 \\
conv\_lstm2d (ConvLSTM2D) & (None, 5, 32, 64, 32)  &   118528 \\
batch\_normalization (BatchNormalization)  & (None, 5, 32, 64, 32)   &  128 \\
spatial\_dropout3d (SpatialDropout3D) & (None, 5, 32, 64, 32) &  0 \\
conv\_lstm2d\_1 (ConvLSTM2D) & (None, 5, 32, 64, 64)   &  614656 \\
batch\_normalization\_1 (BatchNormalization) &  (None, 5, 32, 64, 64)   &  256 \\
spatial\_dropout3d\_1 (SpatialDropout3D)  &   (None, 5, 32, 64, 64)  &   0 \\
conv\_lstm2d\_2 (ConvLSTM2D)  & (None, 5, 32, 64, 128)  &  885248 \\
batch\_normalization\_2 (BatchNormalization)  & (None, 5, 32, 64, 128) &   512 \\
spatial\_dropout3d\_2 (SpatialDropout3D)  &   (None, 5, 32, 64, 128)  &  0 \\
conv\_lstm2d\_3 (ConvLSTM2D)  & (None, 5, 32, 64, 256)  &  394240 \\
batch\_normalization\_3 (BatchNormalization) &  (None, 5, 32, 64, 256)  &  1024 \\
spatial\_dropout3d\_3 (SpatialDropout3D)  &  (None, 5, 32, 64, 256)  &  0 \\
conv2d (Conv2D) &  (None, 5, 32, 64, 5)  &  32005 \\
\hline
Total params: 2,046,597 \\
Trainable params: 2,045,637 \\
Non-trainable params: 960 \\
\end{tabular}}
\end{table} 

Following the standard practice in implementing convolutional models, we also applied a batch normalization layer and a dropout regularization layer with a dropout rate of 0.1 after each convolutional layer (see Table \ref{tab:model} for detailed settings and parameters in each layer). For all \texttt{ConvLSTM} layers, a kernel size of 5 $\times$ 5 and padding \texttt{same} are applied for the first two \texttt{ConvLSTM} layers, and a kernel size of 3 $\times$ 3 and 1$\times$1 are applied to the third and fourth \texttt{ConvLSTM} layers, respectively. 

The AWP model was trained using the root mean square error loss function and the mean absolute error accuracy metric. Due to the large training data, we restrict our batch size to 256 and 100 epochs for the training process. Also, the \texttt{Adam} optimizer with a learning rate of $10^{-3}$ and a decay rate of 0.8 was used, along with a callback early-stopping based on the validation data to save the best model during the training. The whole training of this AWP model on a single A100 GPU took about 5-18 hours, depending on how many channels and/or input data frames used. 

It should be noted that there is no particular formula for designing the AWP model as described above, other than our trials and errors with the given dataset. Our sensitivity experiments with different configurations showed that the AWP model is less sensitive to the number of \texttt{ConvLSTM} layers as long as the number of layers is more than 3 and the number of nodes in the last two layers is not less than 128. The AWP model, however, has worse performance when the kernel size is $>$ 5, which is somewhat similar to what was reported in Shi et al. (2015). As such, we limited the kernel sizes to 1$\times$1, 3$\times$3, and 5$\times$5 for all \texttt{convLSTM} layers in our experiments herein.    

A key step in implementing a deep-learning model based on supervised learning for predicting future state $x_{t+T}$ from a given initial state $x_t$ is to generate data pair $(X,Y) \equiv (x_t,x_{t+T})$ from input data. For example, one could construct these data pairs using two consecutive steps or over a given time window, depending on how the model is used to predict the future state \citep{Shi_etal2015, Nguyen_etal2023, NguyenKieu2024, Kieu2024}. Regardless of how this data pair is constructed, any weather prediction based on recurrent neural networks is very different from the traditional NWP due to the way the component $x_t$ of the pair $(x_t,x_{t+T})$ is constructed and interpreted. Specifically, recurrent models use a range of past information that contains temporal relationships to help improve future prediction, i.e., $x_t$ is now understood as a set of data up to time $t$ ($x_t \equiv [...x_{t-2},x_{t-1},x_t]$). In contrast, a typical NWP model uses a single initial condition slice for future prediction, i.e., $x_t$ is a single time instant at time $t$. To some extent, these recurrent models improve their prediction in the same way that four-dimensional data assimilation optimizes an initial condition over an interval instead of just a single time as for 3D variational data assimilation or optimal interpolation. So the interpretation of data-driven weather models is no longer deterministic. In this regard, the predictability issue is inherent to any data-driven model due to the existence of chaotic characteristics of the atmosphere and random noises in any input data.
%
%
\section{Experiment Description}\label{sec:description}
\subsection{Data}\label{sec:data}
In this study, the reanalysis dataset from the European Center for Medium-Range Weather Forecasting (ERA5) was used to train and validate our AWP model \citep{ERA5}. This ERA5 reanalysis dataset has several different resolution versions \citep{Rasp_etal2020}. For the sake of training and demonstration in this study, we use the global $5^o \times 5^o$ dataset on a latitude-longitude grid, which is given at hourly intervals. For the global coverage, this dataset basically consists of 64$\times$32 data points in the longitudinal/latitudinal directions. With hourly output and 37 pressure levels in the vertical direction, each variable in this ERA5 dataset can be treated as a four-dimensional tensor of the shape $[n_b,32,64,37]$, where $n_b$ is the number of hours between 1979-2018 (i.e., the batch size).     

For our general AWP model development, we used a subset of available variables in ERA5 including relative humidity, three wind components ($u,v,\omega$), geopotential, temperature, total cloud cover, total precipitation, vorticity, 10-meter wind components ($(u_{10},v_{10})$), and landmask. Because these variables have different physical dimensions and vary widely in range and vertical levels, they are all normalized before training, using the maximum/minimum normalization. In addition, only a selected number of vertical levels including 850, 500, and 200 hPa are used instead of all 37 levels for the three-dimensional variables. This restriction is due to our limited memory resources, and so only three levels are used for each variable. 

With 40 years of the ERA5 data at 5-degree resolution, seven three-dimensional variables, five two-dimensional variables and three vertical levels, the input data is a very big 4D tensor of the size $(n_b,n_y,n_x,n_c)$, where $n_b=350640$ is the batch size, $(n_x,n_y) = (64,32)$ is the image dimensions, and $n_c$ is the number of channels varying between [1-26]. To further transform this data into a tensor needed for the LSTM architecture, this dataset is turned into a 5D tensor of dimension $(n_b, n_f, n_y,n_x,n_c)$, where $n_f$ indicates the number of input data frames used to train the ConvLSTM. Following the standard practice in machine learning, we finally split this input data into three subsets for training, validation, and testing with a ratio of 90, 5, and 5\%.  

\subsection{Experiments}\label{sec3.2}
Given our main aim of examining the weather predictability of AI models, the AWP model in Section \ref{sec:model} was therefore designed for the short-to-medium range prediction from an initial time $t=0$ to a later time $t=T$, where $T = 6, 12, 18, 24, 48, 60, 72, 120, 144, 168, 192, 216, 240$ hours are forecast lead times in all experiments. More specifically, we will focus on two different time-stepping techniques for this AWP model, which are based on the small-step and the big-step approaches as follows:

\begin{itemize}
\vskip 0.1in
\item \textit{Small-step approach}: 

In this approach, a short time lag for a data pair $(x_t,x_{t+\Delta})$ is first constructed, where $\Delta$ is the smallest consecutive time interval in the training dataset. This small-step approach is the standard method in current AI models for weather prediction, as it requires one to train only one model and then apply it recursively with time for any lead time. That is, to predict the future state at time $t=T$, one just iterates the model every increment $\Delta$ $n$ times from $t$ to $t+T\equiv t + n\Delta$.

The clear advantage of this small-step approach is that an AI model is guaranteed to be trainable, because $\Delta$ is in general sufficiently small that the time-lag mapping is well maintained or captured in the training data. In general, the smaller the value of $\Delta$, the better the AWP model will be after training. As a result, one can integrate the model up to an arbitrary lead time. So long as the model is still within its predictability limit, a forecast can always be produced. 

Note that while this approach is conceptually very similar to the typical time-stepping in NWP models, we note that the time interval $\Delta$ in our AWP model is practically much bigger than the time step in NWP models due to the enormous data volume, especially for global climate data. Typically, the value of $\Delta$ in most climate datasets varies between 1-6 hours, depending on the spatial resolution of the climate datasets. For our study here, ERA5 is provided at the hourly interval, and so $\Delta=1$ hr. Note also that for the recurrent architectures such as LSTM, the data pair $(x_t,x_{t+\Delta})$ for this small-step approach should be understood in the sense that $x_t$ is a set of input that includes $n_f$ past data slices up to time $t$ at an interval of $\Delta$, i.e., $x_t \equiv \{x_{t-(n_f-1)\Delta}, x_{t-(n_f-2)\Delta},....x_t\}$. As such, the input data for an AWP model includes more information than the initial condition in NWP models. 

\vskip 0.1in
\item \textit{Big-step approach}:

The nature of AI models in learning rules from input data also allows another approach to predict future states at any lead times. Specifically, one can directly construct from input data a set of pair $(X,Y) \equiv (x_t,x_{t+T})$, where $T$ is the actual lead time that one wants to predict. Unlike the small-step approach that propagates from $t$ to $t+T$ by stepping $n$ times using a model for $(X,Y) \equiv (x_t,x_{t+\Delta})$, this approach predicts the future state at time $t+T$ with a single big step whenever the model training is successful, thus justifying its name.

The physical motivation for this big-step approach for AI models is based on the fact that if a weather system is predictable at a lead time $T$, then there must be some relationship between the model state at time $t$ and the future state at time $t+T$, which can be represented mathematically as a propagator from time $t$ to $t+T$ \citep{Kieu2024}. For a physical-based model, this relationship is nothing but the finite numerical scheme that can be formally written as $x_{t+T} = \mathcal{M}(x_t,t,T)$, where the numerical model $\mathcal{M}$ contains all possible parameters and physical representations. For a purely data-driven model, $\mathcal{M}$ is a set of optimal weights obtained by learning rules from input data. 

The clear advantage of this big-step approach is that an AI model will be optimal for a given lead time $T$, so long as there is some relationship between the states at $t$ and $t+T$. For a dynamical system that is deterministic, note that one initial condition at $t$ would give a unique prediction at time $t+T$. So the relationship must exist if we have sufficient training data. Therefore, a good AI model should be able to capture this relationship when there is sufficient data for training, without the need for any governing equation. 

A downside of this approach is that one now has to train a set of models, each model for one specific forecast lead time $T$. In addition, the training may not converge for a long lead time $T$ due to the possible emergence of chaotic behaviors that erase the mapping between two times $t$ and $t+T$ as discussed in \cite{Kieu2024}. Also, this big-step approach will not be applied to hybrid models, which contain both physical-based dynamic components and data-driven physics parameterizations. So, its application is limited within purely data-driven models. 
\end{itemize}. 

It should be mentioned that the big-step approach as outlined above is applied to global AI models only. For regional models, the predictability of any AI model will be also controlled by lateral boundary conditions that the big-step approach is no longer suitable. One way to address this boundary condition issue for regional models is to design an additional input channel that encodes the lateral boundary conditions from global models and then feed this information into a flattened or a dense layer of a regional model. Or, one can treat the lateral boundary as a prompt tuning during the training process. Implementations of these approaches to handle lateral boundary conditions for regional models are however beyond the scope of this study here and so will not be presented herein.  

Given our current computational resource and a large amount of model training for the big-step approach, we present in this report a set of experiments that can be easily run with limited computational resources for quick validation of our results. Specifically, we designed a control experiment with only 5 input channels (i.e., $n_c = 5$) and 5 past data slices (i.e., $n_f=5$). The number of parameters for this control experiment \texttt{Exp1} is given in Table \ref{tab:model}. Along with the control experiments, four sensitivity experiments Exp2-Exp4 were also conducted, with different input channels and data frames. All details of these experiments are summarized in Table \ref{tab:experiments}. A more comprehensive result using more input channels and data frames will be presented in our full article publication.

\begin{table}[ht]
\centering
\caption{Experimental designs for the global AI weather prediction based on the ConvLSTM architecture}
\label{tab:experiments}
\scalebox{1.2}{
\begin{tabular}{cccl}
\hline
Experiments & 
\# channels $n_c$ & 
\# input frames $n_f$ &  
Variables\\
\hline
Exp1 & 5 & 5 & $Z_{500},T_{500},U_{850},V_{850}, RH_{850}$ \\
Exp2 & 2 & 5 & $Z_{500},T_{500}$ \\
Exp3 & 2 & 2 & $Z_{500},T_{500}$\\
Exp4 & 1 & 5 & $Z_{500}$\\
Exp5 & 1 & 2 & $Z_{500}$ \\
\hline
\end{tabular}}
\end{table}
%
%
\section{Results}\label{sec:results}
\subsection{Error growth: big vs small-step approach}
\begin{figure}[ht]
\centering
\includegraphics[width=12cm]{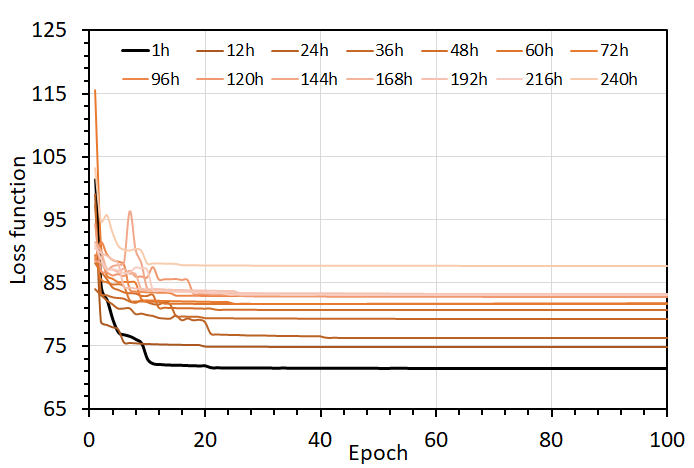}
\caption{Training loss function for different forecast lead time $T = 1, 24, 48, 72, 96, 120, 240$. Note that $T=1$ (black curve) corresponds to the small-step approach, and all other curves (orange spectrum) corresponds to the big-step approach at the corresponding lead time $T$.}
\label{fig1}
\end{figure}

Figure \ref{fig1} shows first the training process for a range of forecast lead times $T=1, 12, 24, 48, 72, 96, 120, 240$ hrs. Note that $T=1$ hr corresponds to the small-step approach (i.e., $T=\Delta$), while all other $T>1$ represent the big-step approach at the corresponding lead times $T$. It is expected to see that the small-step approach indeed allows for the best training (in the sense of its loss function converging towards the smallest value), while it is harder to train the AWP model with the big-step approach for longer lead times. Physically, this behavior simply means that the physical relationship between 2 weather states at times $t$ and $t+T$ becomes quickly deteriorated for longer $T$ due to the limited predictability of the atmosphere. That is, any small difference in the input data at time $t$ could lead to a large difference at a later time $t+T$, making it harder for the AWP model to learn any evolving rules from training data. This explains why the big-step approach converges towards a larger loss value at the end of training for a larger $T$ as shown in Fig. \ref{fig1}, similar to what was obtained from a low-order model in \cite{Kieu2024}.
%
%

\begin{figure}[ht]
\centering
\includegraphics[width=11.5cm]{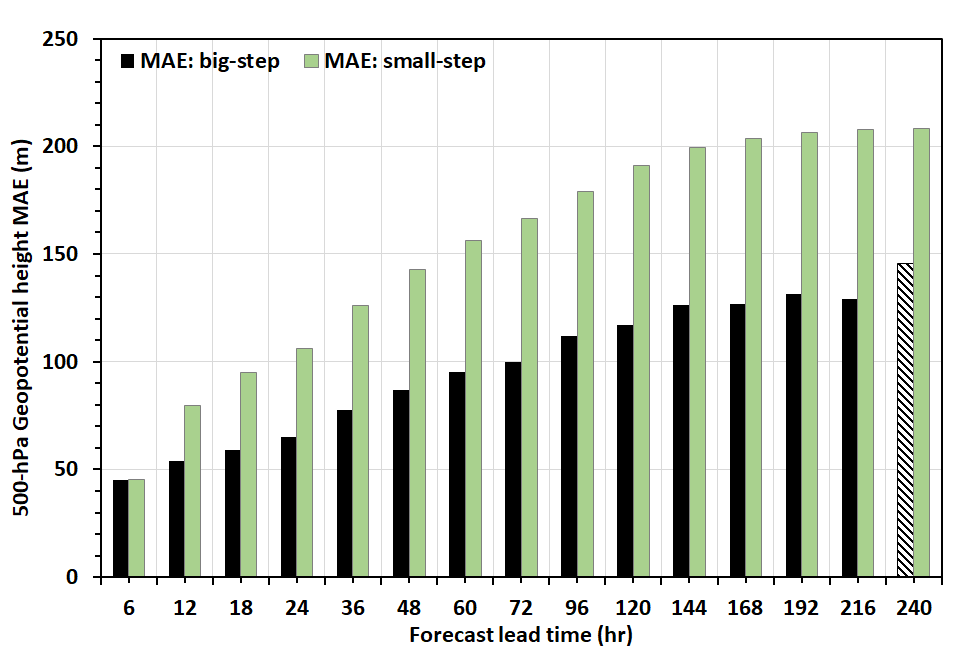}
\caption{The mean absolute errors (MAE) of the 500-hPa geopotential height (columns, unit: m) as a function of forecast lead time (hours) as obtained from the control experiment \texttt{EXp1}, using a test dataset during the example year 2018 for the small-step approach (green) and big-step approach (black). The striped column indicates the difficult training (i.e., slow convergence of the loss function toward a large value shown in Fig. \ref{fig1}) for the lead time $T=240$ hr}
\label{fig2}
\end{figure}

With the successfully trained models, Fig. \ref{fig2} compares the forecast error growth as a function of forecast lead times for 500-hPa geopotential height obtained from the small-step and big-step models, Here, the 500-hPa mean absolute errors (MAE) are computed from an independent test dataset during one illustrative year 2018. It is apparent from these error growth curves that the big-step models noticeably outperform the small-step model at all lead times after just 12 hours, even at the 10-day lead time for which the training of the big-step model could not fully converge (shown as a striped column in Fig. \ref{fig2}). This is an intriguing yet expected result, as it shows that the training for the AWP model at each individual lead time gives better performance than iterating recursively the small-step model. Recall from the above discussion that training an AWP model at the shortest time interval $\Delta$ should give us the best model, because the mapping $\mathcal{M}(x_t,t,\Delta)$ is most preserved. However, using this small-step approach to propagate an initial state $t$ to later time $t+T$ would accumulate errors over time every step, resulting in a quick error growth rate. In contrast, training the AWP model directly between two steps $t$ and $t+T$ gives an optimal model for lead time $T$, thus leading to a lower MAE at all lead times shown in Fig. \ref{fig2}. 

Of further note is that the MAE in both the small-step and big-step models displays an error saturation after 7 days and is significantly larger than the climatological variability of 500-hPa geopotential height, which is $\approx$100 m (\citep{Krishnamurthy2019, Lorenz1982, Weyn_etal2021}). Practically, this implies that the AWP model has a faster error growth rate and a shorter predictability range than physical-based models, whose MAE is saturated around 14 days. Regardless of models or approaches, the existence of error saturation as seen Fig. \ref{fig2} is itself significant, because it indicates the important role of atmospheric limited predictability in weather forecast that no model can overcome after a certain lead time.  

Although using high-resolution training data or more sophisticated AI models may help increase the predictability range or reduce MAE, it is expected that the big-step approach still outperforms the small-step for general time prediction problems due to its design that is optimized at each lead time.   
As seen in Fig. \ref{fig2}, the big-step approach could in fact provide a very comparable MAE relative to climatological forecast, with its 10-day MAE of $\approx$ 120 m despite the same issue with a coarse resolution of training data or a simple model architecture as the small-step approach. In this regard, the significant difference in the MAE between these two approaches highlights the advantage of the big-step approach for predicting weather that we want to present here.   
%
%

\begin{figure*}[!t]%
\centering
\includegraphics[width=17cm]{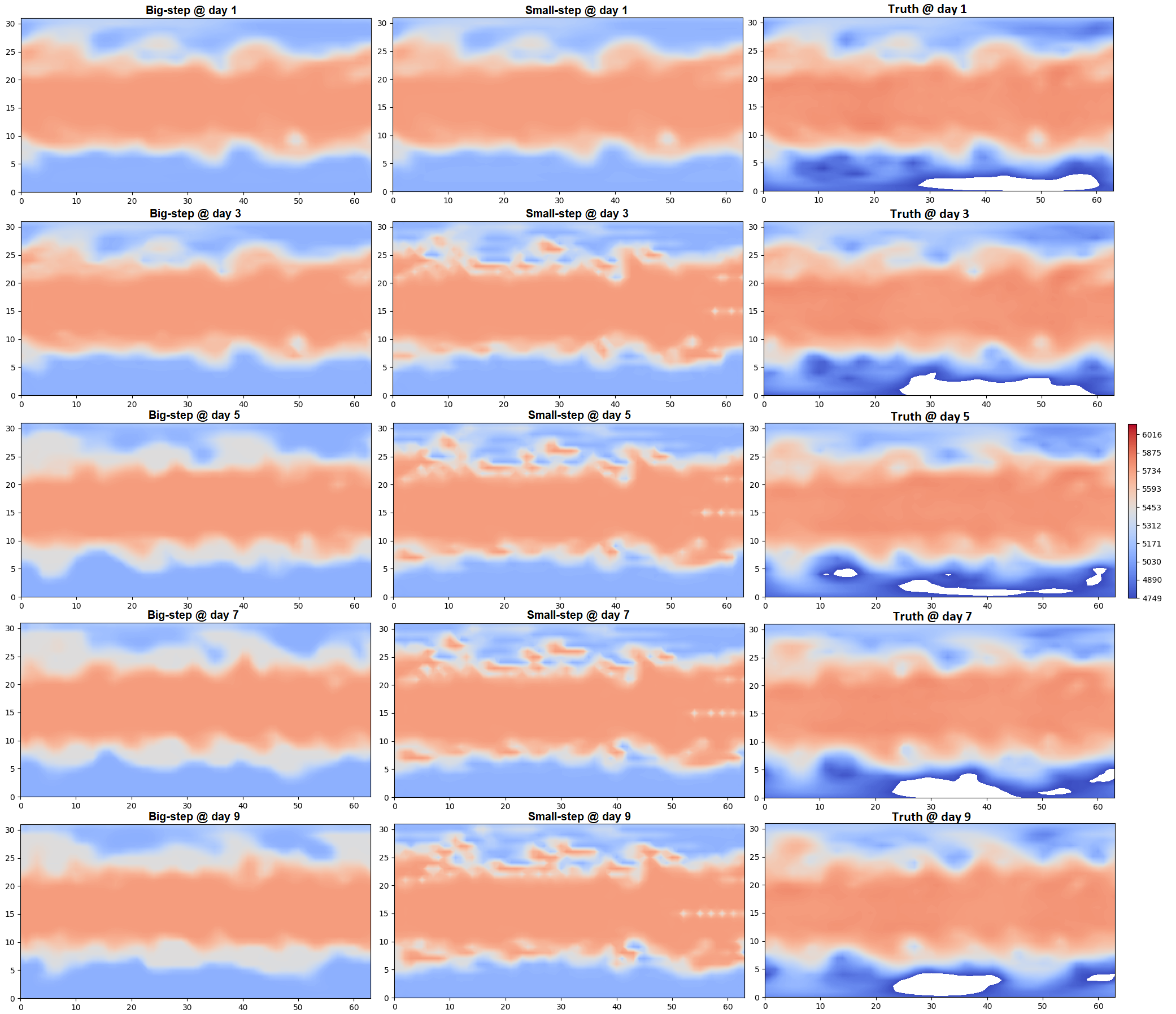}
\caption{An example of global 500-hPa geopotential height predictions (shaded, unit m) initialized at 0000 UTC May 5 2018, which are obtained from the big-step approach (left panels), the small-step approach (middle), and the corresponding validation (i.e., truth) from ERA5 (right panels) at 1, 3, 5, 7, and 9-day forecast lead times.}\label{fig3}
\end{figure*}

We should emphasize that any evaluation of a global model forecast skill based on the 500-hPa geopotential MAE is mostly for an overall evaluation at the synoptic to the global scale. An actual evaluation of a model forecast skill should however depend further on specific weather features that the globally-averaged MAE cannot capture. To illustrate the performance of the small-step and big-step approaches beyond the MAE, Fig. \ref{fig3} shows the distribution of $Z_500$ forecast at 5 different lead times 1,3,5,7 and 9 days, which are verified against the corresponding ERA5 data valid at the same time. While the MAE curve in Fig. \ref{fig2} displays some skill up to 7 days, the actual spatial distribution of predicted $Z_500$ quickly displays some discrepancies from the ERA5 (i.e., ground truth) as soon as 3 days, notably at the polar regions and the extra-tropical transition zones. For example, the ERA5 $Z_500$ at the south pole shows a general dominant low during May, yet both the small-step and big-step approaches could not fully capture this low at all lead times. 

Between the two, the small-step approach shows a sooner breakdown of $Z_500$ anomalies around 30N/S latitude, with some noticeable change in the patterns of $Z_500$ anomaly starting right from day 2 as compared to ERA5 (right panels). Meanwhile, the big-step approach could preserve these high anomalies fairly well until day 7, before they appear to fade away after day 9 when verified against ERA5. Note that in practice it is this type of anomalies that decide day-to-day weather in operational weather forecasts for which the $Z_500$ MAE could not fully represent. From this regard, the AWP model is skillful at the global scale up to 7 days from the MAE perspective, yet it may still not be able to provide good weather forecasts at the regional level beyond 3 days. Similar results are also obtained for other fields such as temperature or horizontal wind (not shown), thus reiterating the different skills based on the MAE or the spatial distribution of a variable. 

Of course, the MAE or the spatial distribution of geopotential height, temperature, or wind anomalies is not the only way to examine the predictability of an AI model. In general, there are many other weather features at different spatial and temporal scales such as MJO, tropical cyclones, squall-lines, mesoscale convective systems, heavy rainfall, etc that require their own metrics when evaluating the forecast skill of a model. These features possess different a predictability range and characteristics that cannot be inferred from $Z_500$ anomalies. In addition to these metric-dependence and feature-specific issues, the predictability of an AI model should be also based on spectral analyses, which can provide a broader picture of how multi-scale error energy spectra grow and reach a saturation limit \citep{KieuRotunno2022, Bonavita2024, Zagar2017}. Such spectral analyses require however much higher resolution data so that high wavenumbers (i.e., smaller spatial scales) can be captured, which are beyond what we present herein. 
Regardless of lacking such spectral analyses for multi-scale systems, the fact that both the MAE curves and spatial patterns of $Z_500$ forecast could capture a consistent property of the AWP model confirms that the big-step approach can provide better forecasts overall at both the regional and global scales. This is an important result, as it suggests one to implement and optimize an AI model for weather prediction at each lead time and scale in operation.     
%
%

\subsection{Sensitivity to input data structure}
An apparent question that we need to address next is how the above behaviors of the AWP model change for different model settings and/or input data types. This question is necessary, as it helps address the robustness of our results on the predictability of the AWP model. To address this question, we first examine in this section a set of experiments in which only 2 input channels including $Z_{500}$ and $T_{500}$ are used for training and prediction instead of all 5 channels shown in Table \ref{tab:experiments}. 

\begin{figure}[ht]
\centering
\includegraphics[width=12cm]{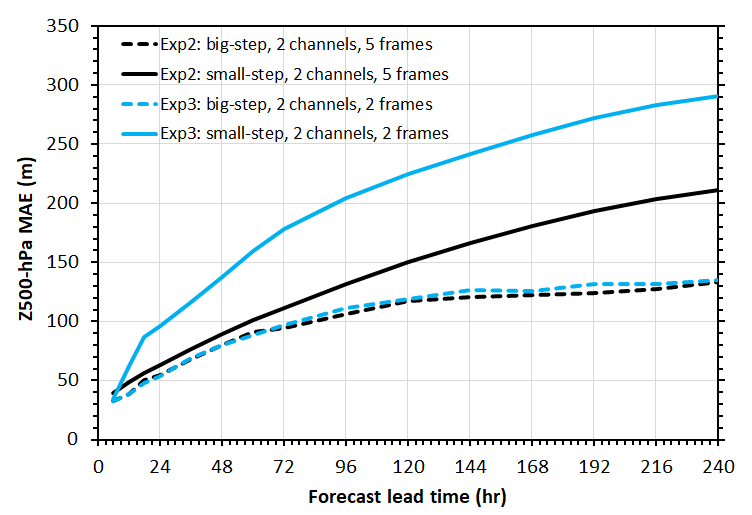}
\caption{Similar to Fig. \ref{fig2}, but for the sensitivity experiments \texttt{Exp2} and \texttt{Exp3} for which only two input channels $Z_{500}, T_{500}$ are used, but using 5 input data frames (\texttt{Exp2}, black) and 2 input data frames (\texttt{Exp3}, blue). Solid and dashed lines denote the small-step and big-step approaches, respectively.}
\label{fig4}
\end{figure}

Figure \ref{fig4} shows the error growth curves for big and small-step approaches as obtained from this 2-channel experiment (\texttt{Exp2}, black lines). It is of interest to note that reducing the number of input channels does not seem to have much of an impact on the overall MAE for both the big-step and small-step approaches, with the MAE only marginally increased at all lead times as compared to the control experiment \texttt{Exp1}. This is very significant, because it indicates that the AWP model can effectively learn the global dynamics of $Z_500$ based mostly on the memory of two key variables $Z_500, T_500$ that removing other channels does not change significantly the overall performance of the model as shown in Fig. \ref{fig4}. 

Reducing the number of input frames (experiment \texttt{Exp3} in Table \ref{tab:experiments}), however, has a substantially bigger impact. As seen in this experiment (blue curves in Fig. \ref{fig4}), fewer input frames lead to an overall much higher MAE, particularly for the small-step approach that shows the rapid growth of the MAE to $\approx$200 m after just 4 days. For the big-step approach, it is surprising to note however that the MAE shows very little increase even with fewer input data frames and channels, albeit still somewhat noticeable. This result clearly suggests that the AWP model could learn well the global dynamics of both $Z_{500}$ and $T_{500}$ with the big-step approach, so long as the training is applied for each individual lead time as designed.      

Intuitively, such an increase in the MAE when reducing the number of input frames is expected, as the model now has less constraint from input data. Thus, the prediction of $Z_500$ can now take a wider range and start to drift away from the mean state more quickly with time. Nevertheless, the better performance of the big-step approach as compared to the small-step approach is still maintained in these experiments regardless of 2 or 5 input frames. Such a robust performance of the big-step approach confirms the effectiveness of this approach in predicting future states, instead of incrementally propagating with time as in the small-step approach.
%
%

\begin{figure}[ht]
\centering
\includegraphics[width=12cm]{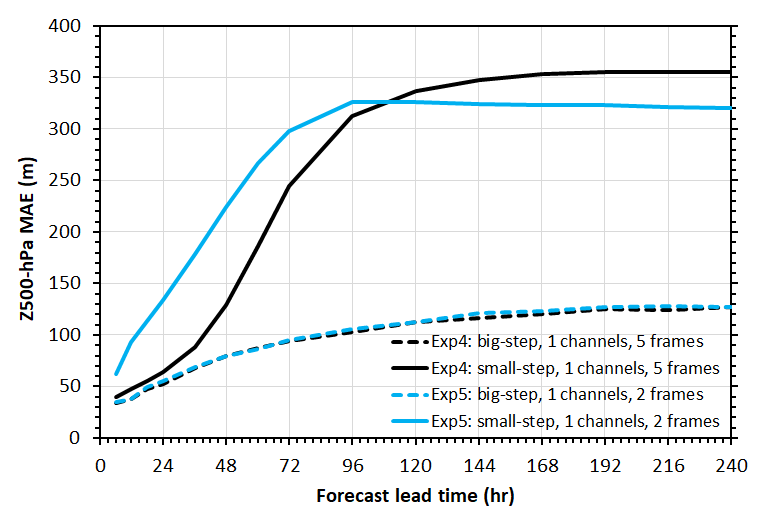}
\caption{Similar to Figure 4, but for the sensitivity experiments \texttt{Exp4} (black) and \texttt{Exp5} (blue) for which only one input channel $Z_{500}$ is used for training and prediction.}
\label{fig5}
\end{figure}

An immediate next question is what if we reduce the number of input channels further down to just one? How does this affect the performance of the small-step and the big-step models at different lead times? In this regard, Fig. \ref{fig5} shows the MAE for another set of sensitivity experiments for which only $Z_{500}$ is used for training, with 2 and 5 input frames similar to Fig. \ref{fig4}. 


The most apparent change in this set of experiments is that the MAE goes up rapidly for the small-step approach, reaching 200 m in less than 2 days and nearly 300 m after just 4 days. Such a rapid degradation of the small-step approach is seen for both $n_f=2$ (\texttt{Exp3}) and $n_f=5$ (\texttt{Exp4}) input frame experiments, revealing the strong impacts of $T_{500}$ on the predictability of $Z_{500}$. Note also that using less number of input frames in experiment \texttt{Exp3} somehow displays worse performance for 0-108 hour lead times than in experiment \texttt{Exp4}, and only shows better performance afterward (see blue curve in Fig. \ref{fig5}. One can further confirm the robustness of these results by trying more input data, different data test periods, or model architectures. However, the key point that using a single channel $Z_{500}$ for input training leads to a faster error growth rate as well as a larger MAE for the small-step approach is expected to remain valid, which will shorten the overall predictability of this approach in AI models. 

We should note that the error growth curves in experiments \texttt{Exp4/Exp5} differ from \texttt{Exp1-Exp3} shown in Fig. \ref{fig4} in one important aspect that the error growth rate (i.e., the slope of the MAE curve) is much slower when using 2 or 5 input channels. Therefore, the predictability range for 2 or 5-channel AWP model is still around 7-9 days before reaching their MAE saturation limit. In contrast, the error growth rate for the 1-channel AWP model is much larger as seen in Fig. \ref{fig5}, which explains a significantly shorter predictability range. From this regard, the strong constraints among some input channels used for training AI models are important when examining the predictability of an AI model.   

Unlike the small-step approach, it is remarkable to note again that the big-step approach is very resilient in terms of the MAE in our AWP model, with almost the same MAE value and predictability range as in the 5 or 2 input channel experiments. Indeed, as seen in Fig. \ref{fig5}, not only the MAE values but also the growth rate of the MAE in experiments \texttt{Exp4/Exp5} is similar to the control experiment \texttt{Exp1}. Because of this, the predictability range in experiments \texttt{Exp4/Exp5} is almost the same as in other experiments for the big-step approach. From a practical perspective, this means our AWP model is still able to predict the future state, even without any other supplementary information from other channels. That is, the global evolution of the atmosphere carries significant memory from the past with or without information from other channels or frames. The fact that the big-step approach performs very consistently in all experiments as seen in Figs. \ref{fig4}--\ref{fig5} reiterates the benefit of this approach for predicting future atmospheric states with AI models. This is an important finding, as it suggests that the development of an AI model for global weather prediction can be sped up significantly with just a few input channels and frames if we adopt the big-step approach for each lead time.   
%
%
%
%
\section{Discussion and Conclusion}\label{sec:conclusion}
In this study, we examined the predictability of AI models for global weather forecasting, using a simple deep-learning architecture based on the \texttt{ConvLSTM}. By implementing two different time-stepping techniques for this AI weather prediction (AWP) model and using ERA5 data for training, we showed first that our AWP model possesses an inherent predictability limit of 7-9 days as expected, albeit it is relatively shorter than the current physical-based model. Such a short predictability range in our AWP model is likely attributed to the coarse resolution of training data as well as our model architecture that may not be optimal for weather prediction. 

Despite these potential shortcomings, the results obtained in this study suffice to highlight several important predictability issues in AI models that we wish to present in this report. First, the limited predictability as obtained in our AWP model reiterates that weather systems contain a chaotic nature that prevents one from predicting weather indefinitely over time, regardless of using physical-based or data-driven models. In fact, this predictability issue seems to be more severe for data-driven models in the sense that these models have to learn physical/dynamical rules from input data, most often few limited sets of climate reanalysis or observations. Because these datasets will always contain uncertainty and cover only a finite phase space of atmospheric dynamics, data-driven models will lose their ability to learn rules as soon as a weather system enters its chaotic regime. One can always train an AI model with a given accuracy threshold, but the model will eventually no longer be able to distinguish the future states of two almost identical inputs (within a noise level of input data) after some lead time. Any good AI model must reflect this fact, which cannot be eliminated due to our incomplete understanding and/or observation of atmospheric states in any meteorological dataset.   

Second, unlike NWP models that contain the predictability of various systems across spatial and temporal scales as coherent and intrinsic properties of the governing equations, the predictability of an AI model must be customized for each metric in practical applications. That is, there is no single AI model that can be used for all general weather prediction purposes at different scales as NWP models. This is because each AI model is trained for a specific purpose, data type, region, and accuracy threshold, which cannot be however applied to all spatial and temporal scales of weather systems. In an ideal scenario when one has an infinite amount of observational data and computational resources for training an AI model, the model can be best optimized for one specific accuracy metric such as the global average absolute errors of geopotential height, temperature, wind,... However, this model is still good only for that data resolution and metrics, and it may not be optimal for temporal/spatial scales unresolved by the input data or different accuracy metrics such as tornado frequency, tropical cyclone intensity, or rainfall rate for which the mean error of geopotential height or temperature alone cannot capture.  

Of course, one may further attempt to construct a loss function that encompasses all possible metrics and scales for the training of a perfect AI model, with its loss function reaching to zero at every lead time. However, note that such a perfect model will not be able to eliminate the fact that atmospheric processes are chaotic and contain some unavoidable uncertainty at any state. Thus, a small noise unresolved in any test/validation input data will lead to different future states after some lead times for which the AI model must capture similar to NWP models, no matter how perfect the AI model is. As a result, such an AI model will display limited predictability for any weather feature, accuracy metric, forecast lead time, and region that one designs it for. From this perspective, one has to develop different AI models for different weather purposes, instead of a unified modeling framework for a typical NWP model in operational centers.        

Using two different time-stepping approaches for our AWP model, it was also found in this study that the big-step approach, which directly predicts the future state at lead time $t=T$ from an initial time $t=0$, outperforms the common small-step approach in current AI models, which recursively iterates every small time increment from $t=0$ to $t=T$. This big-step approach is very effective and resilient to different model input channels and frames, with a consistent 500-hPa geopotential height MAE of 120 m at 7-10 days. Such outperformance of the big-step approach is expected, as the training for the AWP model is optimized at each lead time $T$. Thus, it suggests a good way to implement AI models for practical weather prediction.  

In contrast, the small-step approach is very sensitive to the number of input channels or input data frames. For 5 input channels and 5 input data frames, it reaches a saturated MAE value of 200 m after 7 days. However, its MAE saturates quickly after just 3 days at a limit of $\approx$300 m when 1 input channel and 1 input frame are used. Of course, these results were obtained for our specific AWP model based on the ConvLSTM architecture. One could improve this AWP model by fine-tuning different model hyperparameters, introducing more layers, adding some attention mechanisms, or using high-resolution ERA5 to help improve the model performance. However, the fact that the big-step approach could reach a comparable climatological level of $Z_500$ variability, as well as its persistence in forecast skill across data types, frames, and lead times even with a minimal model and coarse data resolution strongly supports its advantages relative to the small-step approach. 

As a final note, we wish to mention that a good AI model will generally converge during training so long as there exists some rules in the training data and the model is properly designed. However, as soon as the AI model is successfully trained (or pre-trained), it immediately implies that the model would not be suitable for prediction in a chaotic regime, because by definition, AI models should not be trainable with the big-step approach in that chaotic regime. This opens up a way to quantify the predictability range in an AI model based on the big-step approach as proposed in \cite{Kieu2024}. Specifically, one can search for a lead time at which the training of the AI model is no longer feasible (i.e., the loss function does not converge with epochs). Such a range is guaranteed to exist due to the chaotic nature of the atmosphere, and so a predictability range for the AI model can always be obtained for any weather features, input data types, or model architecture. More results on this application for other weather features such as heavy rainfall  will be presented in our upcoming studies. 


\section*{Acknowledgments}
This research was funded by the NSF (AGS \# 2309929).

\bibliographystyle{unsrtnat}

\end{document}